\documentclass[10pt]{article}

\usepackage{amsmath}
\usepackage{amssymb}
\usepackage{graphics}
\usepackage{rotating}
\usepackage{cite}
\usepackage{color}
\usepackage{fancybox}
\usepackage{pstricks}
\usepackage{xcolor}
\usepackage{multicol}

\def\0{\mbox{\tiny $0$}}
\def\1{\mbox{\tiny $1$}}
\def\2{\mbox{\tiny $2$}}
\def\3{\mbox{\tiny $3$}}
\def\4{\mbox{\tiny $4$}}
\def\5{\mbox{\tiny $5$}}
\def\6{\mbox{\tiny $6$}}
\def\7{\mbox{\tiny $7$}}
\def\8{\mbox{\tiny $8$}}
\def\9{\mbox{\tiny $9$}}
\definecolor{navy}{rgb}{0,0,.6}
\definecolor{jour}{rgb}{0,0.6,.4}
\definecolor{jbul}{rgb}{0.7,0.,.4}
\begin{document}
%
\thispagestyle{empty}
\setcounter{page}{0}

\begin{center}
{\colorbox{gray!15}{
{\color{navy} \bf \Large
\begin{tabular}{c}
AXIAL DEPENDENCE OF OPTICAL WEAK \\
MEASUREMENTS IN THE CRITICAL REGION
 \end{tabular}
}
}}
\end{center}

\vspace*{1cm}

\begin{center}
{\large
{\color{jbul}
$\boldsymbol{\bullet}$}
{\color{jour}
\bf Journal of Optics 17, 035608-10 (2015)}
{\color{jbul}
$\boldsymbol{\bullet}$}}
\end{center}

\vspace*{1cm}

\begin{center}
\begin{tabular}{cc}
\begin{minipage}[t]{0,5\textwidth}
{\bf Abstract}.
The interference between optical beams of different polarizations plays a fundamental role in reproducing the optical analog of the electron spin {\em weak measurement}. The extraordinary point in {\em optical} weak measurements  is  represented by the possibility to estimate with great accuracy  the {\em Goos-H\"anchen (GH) shift}\, by measuring the distance between the peak of the outgoing beams for  two opposite rotation angles of the polarizers located before and after the dielectric block. Starting from the numerical calculation of the  GH shift, which clearly shows a frequency crossover for incidence  near to the critical angle,  we present a detailed study of the interference  between {\em s} and {\em p} polarized waves in the critical region. This allows to determine in which conditions it is possible to avoid axial deformations and  reproduce the GH curves.
In view of a possible experimental implementation,  we give the {\em expected}\, weak measurement curves  for gaussian lasers of different beam waist sizes  propagating through borosilicate (BK7) and fused silica dielectric blocks.
\end{minipage}
& {\colorbox{gray!15!}{
\begin{minipage}[t]{0,4\textwidth}
{\bf Manoel P. Ara\'ujo}\\
Institute of  Physics ``Gleb Wataghin''\\
State University of Campinas (Brazil)\\
{\color{navy}{{\bf mparaujo@ifi.unicamp.br}}}
\hrule
\vspace*{0.1cm}
{\bf Stefano De Leo}\\
Department of Applied Mathematics\\
State University of Campinas (Brazil)\\
{\color{navy}{{\bf deleo@ime.unicamp.br}}}
\hrule
\vspace*{0.1cm}
{\bf Gabriel G. Maia}\\
Institute of  Physics ``Gleb Wataghin''  \\
State University of Campinas (Brazil)\\
{\color{navy}{{\bf ggm11@ifi.unicamp.br}}}
\end{minipage}
}}
\end{tabular}
\end{center}

\vspace*{1cm}

\begin{center}
{\color{navy}
{\bf
\begin{tabular}{ll}
I. & INTRODUCTION \\
II. & THE GH SHIFT FOR BK7 AND FUSED SILICA BLOCKS \\
III.  & WEAK MEASUREMENTS IN OPTICAL EXPERIMENTS \\
IV. & THE PEAKS BEHAVIOR IN THE CRITICAL ANGLE REGION\\
V. & CONCLUSIONS AND OUTLOOKS\\
& \\
& \,[\,16 pages, 7 figures\,]
\end{tabular}
}}
\end{center}

\vspace*{3.75cm}

{\large
\begin{flushright}
{\color{jbul}
$\boldsymbol{\bullet}$}
{\color{jour}
$\boldsymbol{\Sigma\hspace*{0.06cm}\delta\hspace*{0.035cm}\Lambda}$}
{\color{jbul}
$\boldsymbol{\bullet}$}
\end{flushright}
}


\newpage


\section*{\large \color{navy} I. INTRODUCTION}

The Goos-H\"anchen [GH] shift\cite{R1a,R1b,R1c} surely represents one of the most intriguing research subjects appeared in literature in the last decades\cite{R2a,R2b,R2c,R2d,R2e,R2f,R2g,R2h,R2i}. This shift, which is  probably one of the clearest manifestations of the evanescent nature of light, represents an additional contribution to the geometrical optical path predicted by the Snell law\cite{R3a,R3b}. This quantum effect  is  still subject of careful and broad investigation and continues to stimulate new discussions\cite{R4a,R4b,R4c,R4d,R4e,R4f,R4g,R4h}. Of particular interest tothe study presented in this paper, it is  the GH shift frequency crossover\cite{R5}.  For  incidence angles, $\theta_{\0}$,  far from  the critical angle, $\theta_c$,  it is well know that the GH shift  is proportional to the wavelength, $\lambda$, of the optical beam\cite{R1b,R2a,R5,R6}.  For incidence   at {\em critical angle} the GH shift is amplified by a factor $\sqrt{{\rm w}_{\0}/\lambda}$, where ${\rm w}_{\0}$ is the beam waist.  This amplification  has been recently obtained  analytically  by using the {\em stationary phase method} \cite{spm1,spm2} and then confirmed by numerical calculations\cite{R5}. This frequency crossover will play a fundamental role in  deriving the {\em expected}  experimental curves for {\em optical} weak measurements in the critical (angle) region.

In a recent interesting experimental paper\cite{R7}, by using the optical analog\cite{R8a,R8b,R8c} of the electron spin weak measurement\cite{R9},  the behavior of the GH shift curve  has been reproduced in the region in which  the incidence angles are {\em far enough} from the critical angle to permit some important approximations.

It is important to observe that in the optical analog of the electron spin weak measurement, polarized light plays the role of the spin $\frac{1}{2}$ particles and the laser beam replaces the coherent electron beam. Due to the fact that the displacement produced by the optical system is a lateral shift rather than an angular deflection, as happens in presence of the Stern-Gerlach magnet\cite{R9}, we have to consider spatial distributions instead of momentum distributions. Notwithstanding  the physics is far from being the same, by using  the needed attention, an optical version of the electron spin weak measurement experiment can be constructed\cite{R8a}.

A unified linear algebra approach to dielectric reflection, recently appeared in litterature\cite{R8b,R8c}, bases the analogy between weak values and optical beam shifts of polarized waves on the {\em expectation value} of the Artman operator. Such an operator is shown to be Hermitian for total internal reflection and non-Hermitian in the critical region\cite{R8b}. For the mathematical details, we refer the reader to ref.\cite{R8c}. In our approach, we discuss the optical analog of the electron spin weak measurement, by analyzing, as done theoretically  in ref.\cite{R8a} and experimentally in ref.\cite{R7},  the distance between the {\em peaks} of the outgoing optical beam. We recall, that in the critical region, due to the breaking of symmetry\cite{R4h}, peak and mean value does {\em not} necessarily coincide. In view of these comments, our discussion can be seen as a complementary work to that one which appears in refs.\cite{R8b,R8c}.

To make this introduction and  the objective of our analysis clearer to the reader, we recall that
in the optical analog of the quantum weak measurement\cite{R7,R8a}, the  parameters which characterize the behavior of the peaks distance in the experimental curves  are
\begin{equation}
\epsilon =  \epsilon_{\0} + \Delta \epsilon =  \cos(\alpha-\beta)/\cos(\alpha+\beta)   \,\,,
 \end{equation}
where  $\alpha$ and $\beta=\alpha+\frac{\pi}{2}\,+\,\gamma_{\0}\,+\,\Delta \gamma$  are the polarization angles  of the first and second polarizer, see Fig.\,1,  and
\begin{equation}
\Delta y_{_{\rm GH}}=y_{_{\rm GH}}^{^{[p]}} - y_{_{\rm GH}}^{^{[s]}}\,\,,
\end{equation}
where $y_{_{\rm GH}}^{^{[s,p]}}$  are respectively the GH shifts for $s$ and $p$  polarization. For small rotation angles, i.e.  $\Delta \gamma \ll 1$, and for an incoming beam with an equal mixture of polarizations, i.e.
$\alpha=\pi/4$, we have
\[
\epsilon_{\0}+\Delta \epsilon = \tan(\gamma_{\0}+\Delta\gamma)\approx \tan\gamma_{\0} + \frac{\Delta \gamma}{\cos^{^{2}}\gamma_{\0}}\,\,.
\]
For incidence angles far from the critical angle, the condition
\begin{equation}
\label{gamma}
 \Delta \epsilon \gg  \Delta y_{_{\rm GH}}/{\rm w}(z) \approx \lambda/{\rm w}(z)\,\,,
 \end{equation}
 where ${\rm w}(z)={\rm w}_{\0}\sqrt{1+\left(\lambda\,z/\pi\,{\rm w}_{\0}^{^{2}}\right)^{^{2}}} $,
is easily satisfied  and, as we shall see in detail later, the distance between the peak  of the outgoing beams  for two opposite rotations in the second polarizer, i.e. $\beta_{_{\pm}}= \frac{3}{4}\,\pi+\gamma_{\0}\pm |\Delta \gamma|$,  is given by
\begin{equation}
\Delta Y_{_{\rm max}} \approx   \,\, \Delta y_{_{\rm GH}}/|\Delta \epsilon|\,\,.
\end{equation}
 Consequently, for polarizer rotations which satisfy the constraint (\ref{gamma}),
 the experimental curve of $\Delta Y_{_{\rm max}}$ reproduces the  GH  shift curve {\em amplified} by the factor $1/|\Delta \epsilon|$. The GH behavior and its amplification (far from the critical region) has been recently confirmed in the experimental investigation presented in ref.\,\cite{R7}.

As observed in the begin of this introduction, the frequency crossover in the critical region\cite{R5} leads to
\begin{equation}
\Delta y^{^{\rm [cri]}}_{_{\rm GH}} \,\propto\,  \sqrt{\lambda\,{\rm w}(z)}\,\,\gg \,\lambda  \,\,.
\end{equation}
This critical GH shift behavior stimulates the investigation of what happens for incidence angles  near to the critical angle, where, due to the amplification $\sqrt{{\rm w}(z)/\lambda}$,   the condition (\ref{gamma}) could be {\em no} longer valid. This should modify the shape of the experimental curves and a {\em new} formula should be introduced to estimate the GH shift by the measurement of the peaks distance, $\Delta Y_{_{\rm max}}$.
In view of a possible experimental implementation of optical weak measurements for incidence {\em near} to the critical angle, we analyze the expected experimental curves for  mixed  polarized laser gaussian beams, with $\lambda=633\,\,{\rm nm}$  and  $\mbox{w}_{\0}=200,\,300,\,500\,\,\mu{\rm m}$,  propagating through BK7 and fused silica dielectric blocks.

The paper is organized as follows. In section II, we give the transmission coefficient for the beam propagating through the dielectric structure of Fig.\,2 and calculate the axial dependence of the GH shift for BK7 (Fig.\,3) and fused silica (Fig.\,4) blocks. In section III, we introduce the idea of weak measurement in optics and analyze the effect that polarizer and analyzer  play  on the  $s$ and $p$ polarized waves of the outgoing beam. For critical incidence, new parameters  have to be introduced (Fig.\,5). The analysis of the distance between the main peak of the outgoing beams for two opposite rotation angles of the second polarizer is presented in section IV. There a {\em new} analytical relation between the GH shift and the peaks distance is also introduced  In this section, we  also present the {\em expected}\,\,experimental curves for incoming gaussian beams with different beam waists (${\rm w}_{\0}=200,\,300,\,500\,\mu{\rm m}$) propagating through BK7 (Fig.\,6) and fused silica (Fig.\,7) dielectric blocks.
The axial dependence of optical weak measurements is clear in the plots and represents one of the important results of our analysis. Conclusions and outlooks  are drawn in the final section.

\section*{\large \color{navy} II. THE GH SHIFT FOR BK7 AND FUSED SILICA BLOCKS}

In order to obtain the mathematical expression for the transmitted beam, $E_{_{\rm out}}^{^{[s,p]}}$, propagating in the $y$-$z$ plane through the dielectric block (see Fig.\,1 and Fig.\,2),  let us first introduce the gaussian wave number distribution which determines  the shape of the incoming beam, $E_{_{\rm in}}^{^{[s,p]}}$,
\begin{equation}
g(\theta-\theta_{\0})= \frac{k\,\mbox{w}_{\0}}{2 \,\sqrt{\pi}}\, \exp \left[-\,(\,k\,\mbox{w}_{\0}\,)^{^{2}} (\theta -\theta_{\0})^{^2}/\,4\,\right]\,\,.
\end{equation}
In the electric amplitude expressions the superscript notation distinguishes between $s$ and $p$ polarized light.
By using the paraxial approximation ($k\,{\rm w}_{\0} \gtrsim 10$),  the incoming electric field, which
moves from the source laser to the left side of the dielectric block, can be represented by\cite{R3a,R3b}
\begin{eqnarray}
E_{_{\rm in}}^{^{[s,p]}}(y,z)
& = &  E_{\0} \,e^{ik\,z}\,\int_{{-\,\pi/2}}^{{+\,\pi/2}}\hspace*{-.7cm}
\mbox{d}\theta\,\, g(\theta-\theta_{\0})\,  \, \exp
\left[\,i\,\left(\theta-\theta_{\0}\right)\,k\,y  -\,i\ \frac{ (\theta-\theta_{\0})^{^2}}{2}\,\,k\,z \right]\nonumber\\
 & = & \frac{E_{\0}\,\,e^{ik\,z}}{\sqrt{1+2\,i\,
 \displaystyle{\frac{z}{k\,\mbox{w}_{\0}^{\2}}}}}\,\,\exp \left[-\frac{
y^{\2}}{\mbox{w}_{\0}^{\2}+2\,i\,\displaystyle{\frac{z}{k} }}\,\right]\,\,.
\label{Ein}
\end{eqnarray}
For gaussian lasers with a small beam waist  with respect to the
dimensions of the dielectric block, we can use the step technique of quantum mechanics\cite{del1,del2,del3,del4,del5} and give the Fresnel coefficients in terms of the angle $\theta$, $\psi$, and $\varphi$ (see Fig.\,2),
\[\sin\theta = n\,\sin \psi\hspace*{1cm}{\rm and}\hspace*{1cm}\varphi=\psi+\frac{\pi}{4}\,\,.\]
The  transmission coefficient which characterizes the outgoing beam is obtained by the transmission through
the left/right sides and the reflection between the up/down sides  of the dielectric block. After simple algebraic manipulations (for more details see ref.\cite{del6}), we find
 \begin{equation}
T^{^{[s]}}(\theta) =\frac{4\,n\cos\psi \cos \theta}{\left(\,\cos\theta + n\cos \psi\,\right)^{^{2}}}
\left(\,\frac{n\cos \varphi - \sqrt{1-n^{^{2}}\sin^{\2}\varphi}} {n\cos \varphi + \sqrt{1-n^{^{2}}\sin^{\2}\varphi}}
\,\right)^{^{2}}\,\,
\exp[\,i\,\phi_{_{\rm Snell}}\,]
\label{TraS}
\end{equation}
and
 \begin{equation}
 T^{^{[p]}}(\theta)= \frac{4\,n\cos\psi \cos \theta}{\left(\,n\cos\theta + \cos \psi\,\right)^{^{2}}}
\left(\,\frac{\cos \varphi - n\sqrt{1-n^{^{2}}\sin^{\2}\varphi}} {\cos \varphi + n\sqrt{1-n^{^{2}}\sin^{\2}\varphi}}
\,\right)^{^{2}}\,\,
\exp[\,i\,\phi_{_{\rm Snell}}\,]\,\,,
\label{TraP}
\end{equation}
where
\[ \phi_{_{\rm Snell}}=  k\,\left[\,
\sqrt{2}\,n\cos\varphi \, \overline{AB} +  (n\cos\psi- \cos\theta)\, \frac{\overline{AD}}{\sqrt{2}}\,\right]\,\,.\]
For $n\sin\varphi<1$,  the outgoing beam,
\begin{equation}
E_{_{\rm out}}^{^{[s,p]}}(y,z)
 =   E_{\0} \,e^{ik\,z}\,\int_{{-\,\pi/2}}^{{+\,\pi/2}}\hspace*{-.7cm}
\mbox{d}\theta\,\, T^{^{[s,p]}}(\theta)\,g(\theta-\theta_{\0})\,  \, \exp
\left[\,i\,\left(\theta-\theta_{\0}\right)\,k\,y  -\,i\ \frac{ (\theta-\theta_{\0})^{^2}}{2}\,\,k\,z \right]\,\,,
\label{Eout}
\end{equation}
is centered at
\begin{eqnarray}
\label{SnellP}
y_{_{\rm Snell}}=\,-\, \left[\,\frac{\partial\phi_{_{\rm Snell}}}{k\,\partial \theta}\,\right]_{_0} =
\, \cos\theta_{\0}\, \left[\, (\,\tan\psi_{\0} + 1\,)\,\overline{AB}  + (\,\tan\psi_{\0} - \tan\theta_{\0}\,)\, \frac{\overline{AD}}{\sqrt{2}} \,\right]\,\,.
\end{eqnarray}
It represents the well-known geometrical shift predicted by the Snell law in ray optics.

For $n\sin\varphi>1$ an additional phase comes from the internal reflection
coefficients in (\ref{TraS}) and (\ref{TraP}),
\begin{equation}
\left\{\,\phi_{_{\rm GH}}^{^{\,[s]}}\,,\,\phi_{_{\rm GH}}^{^{\,[p]}}
\,\right\}
= -\,4\,\left\{\,\arctan\left[\,\frac{\sqrt{n^{\2}\sin^{\2}\varphi-1}}{n\cos\varphi}\,\right]\,,\,
\arctan\left[\,\frac{n\,\sqrt{n^{\2}\sin^{\2}\varphi-1}}{\cos\varphi}\,\right]
\,\right\}
\end{equation}
and a new  shift  (the well known GH shift)  has to be considered. The numerical data for the propagation
through  BK7 ($n=1.515$) and fused silica ($n=1.457$) are respectively plotted in Fig.\,3 and Fig.\,4. The data  clearly show the amplification for incidence in the critical region and they are in  excellent agreement with the analytical prediction for incidence far from the critical angle,
\begin{eqnarray}
\label{anaGH}
\left\{\,y_{_{\rm GH}}^{^{[s]}}\,,\,y_{_{\rm GH}}^{^{[p]}}\,\right\} &=& - \left[\,
\left\{\,\frac{\partial \phi_{_{\rm GH}}^{^{\,[s]}}}{k\,\partial \theta}\,,\,
\frac{\partial \phi_{_{\rm GH}}^{^{\,[p]}}}{k\,\partial \theta}
\,\right\}\,\right]_{_0} \nonumber \\
 &=& \frac{4\,\cos\theta_{\0}\,\sin\varphi_{\0}}{k\,\cos\psi_{\0}\,\sqrt{n^{^{2}}\sin^{\2}\varphi_{\0}-1}}\,
 \left\{\,1\,,\,\frac{1}{n^{^{2}}\sin^{\2}\varphi_{\0} - \cos^{\2}\varphi_{\0}}\,\right\}\,\,,
\end{eqnarray}
where the ${\rm w}_{\0}$ dependence disappears\cite{R1b,R2a,R5,R6}. The plots of the GH shift for BK7 and fused silica clearly show an {\em axial} dependence. This axial dependence, which  has been recently investigated and recognized as a possible source of  angular deviations in the optical path predicted by the Snell law\cite{del6}, has to be  seen in the optical weak measurement curves as well. Understanding how this axial dependence modifies the optical weak measurement curves in the critical region is one of the main objectives of our study.

\section*{\large \color{navy} III. WEAK MEASUREMENTS IN OPTICAL EXPERIMENTS}

On the basis of the results presented in the previous section, we can approximate the $s$ and $p$ polarized outgoing beams as follows
\begin{equation}
E_{_{\rm out}}^{^{[s,p]}}(y,z) \approx  \frac{E_{\0}\,\,e^{ik\,z}}{\sqrt{1+2\,i\,
 \displaystyle{\frac{z}{k\,\mbox{w}_{\0}^{\2}}}}}\,\,\left| T^{^{[s,p]}}_{\0}\right|\,\,\exp \left[-\frac{
\left(y-y_{_{\rm Snell}} -y_{_{\rm GH}}^{^{[s,p]}}  \right)^{\2}}{\mbox{w}_{\0}^{\2}+2\,i\,\displaystyle{\frac{z}{k\,}} } + i\,\left( \phi_{_{\rm Snell,0}}+  \phi_{_{\rm GH,\0}}^{^{[s,p]}}\right)\,\right]\,\,,
\end{equation}
where for  $y_{_{\rm GH}}^{^{[s,p]}}$, which represents the only entry for which we have not a full analytical expression,  we have to use the numerical data plotted in Fig.\,3 and Fig.\,4.

The intensity of the outgoing beam coming out from the dielectric block and moving  towards the analyzer ($z_{_{\rm out}}<z<z_{_{\rm A}}$ in Fig.\,1) is given by
\begin{eqnarray}
I_{_{\rm out}}(y,z_{_{\rm out}}<z<z_{_{\rm A}})  & = & \left|\, \sin \alpha \, E^{^{[s]}}_{_{\rm out}}(y,z)   +    \cos \alpha \, E^{^{[p]}}_{_{\rm out}}(y,z)   \right|^{^{2}}\nonumber \\
 & & \hspace*{-4.8cm}\propto \,\,
  \left|\,\tau\,\tan\alpha\,\exp\left[-\,\left(\,\frac{y-y_{_{\rm Snell}}-y^{^{[s]}}_{_{\rm GH}}}{{\rm w}(z)}\right)^{^{2}} + \,i\,\Delta \phi_{_{\rm GH}} \,\right]  +\, \exp\left[-\,\left(\,\frac{y-y_{_{\rm Snell}}-y^{^{[p]}}_{_{\rm GH}}}{{\rm w}(z)}\right)^{^{2}} \,\right]     \right|^{^{2}},
\end{eqnarray}
where $\tau = \left|\,T^{^{[s]}}_{\0}/\,\,T^{^{[p]}}_{\0}\right|$ and $\Delta \phi_{_{\rm GH}}= \phi^{^{[s]}}_{_{\rm GH,\0}}-\phi^{^{[p]}}_{_{\rm GH,\0}}$. The $\theta_{\0}$ dependence of  $\tau$  and $\Delta \phi_{_{\rm GH}}$ is plotted in Fig.\,5a (BK7) and Fig.\,5b (fused silica). After removing the phase difference between the $s$ and $p$ polarized light by the analyzer located at $z=z_{_{\rm A}}$ and combining $s$ and $p$ polarization by the second polarizer located at $z=z_{_{\beta}}$, the outgoing beam intensity becomes
\begin{equation}
I_{_{\rm out}}(Y,z>z_{_\beta}) \,\propto \,\,
  \left|\,\tau\,\,\tan\alpha\,\tan\beta\,\,\exp\left[-\,\left(\frac{Y +\,\displaystyle{\frac{\Delta y_{_{\rm GH}}}{2}}}{{\rm w}(z)}
  \right)^{^{2}} \right]  +\, \exp\left[-\,\left(\frac{Y -\,\displaystyle{\frac{\Delta y_{_{\rm GH}}}{2}}}{{\rm w}(z)}\right)^{^{2}} \,\right]     \right|^{^{2}},
\end{equation}
where
\[
Y  =  y \,-\,  y_{_{\rm Snell}}\,-\,\frac{ y^{^{[s]}}_{_{\rm GH}}+y^{^{[p]}}_{_{\rm GH}}}{2}
\,\,\,\,\,\,\,\,\,\,\,
{\rm and}
\,\,\,\,\,\,\,\,\,\,\,
\Delta y_{_{\rm GH}}=y^{^{[p]}}_{_{\rm GH}}-\,y^{^{[s]}}_{_{\rm GH}}\,\,.
\]
Observing that
\[
\tan\alpha\,\tan\beta \,=\, \frac{\epsilon-1}{\epsilon+1} \,\approx\,
\frac{\epsilon_{\0}-1}{\epsilon_{\0}+1} + \frac{2}{(\epsilon_{\0}+1)^{^{2}}}\,\,\Delta \epsilon\,\,,
\]
the choice of an appropriate rotation $\gamma_{\0}$ permits to fix the parameter $\epsilon_{\0}(=\tan\gamma_{\0})$ to
\begin{equation}
\epsilon_{\0}=\frac{\tau -1}{1+\tau}\,\,\,\,\,\,\,\Rightarrow\,\,\,\,\,\,\,\frac{\epsilon_{\0}-1}{\epsilon_{\0}+1} = -\,\frac{1}{\tau}\,\,.
\end{equation}
The angular dependence of $\gamma_{\0}$ is plotted in Fig.\,5b (BK7) and Fig.\,5d (fused silica) for incidence angle in the critical region. This choice allows to rewritten  the outgoing intensity in terms of the parameter $\tau$ and $\Delta \epsilon$ as follows
\begin{eqnarray}
I_{_{\rm out}}(Y,\Delta \epsilon) & \propto &
  \left|\,\left[\, \frac{(1+\tau)^{^{2}}}{2\,\tau}\,\Delta\epsilon\,-1 \,\right]\,\exp\left[-\,\left(\frac{Y +\,\displaystyle{\frac{\Delta y_{_{\rm GH}}}{2}}}{{\rm w}(z)}
  \right)^{^{2}} \right]  +\, \exp\left[-\,\left(\frac{Y -\,\displaystyle{\frac{\Delta y_{_{\rm GH}}}{2}}}{{\rm w}(z)}\right)^{^{2}} \,\right]     \right|^{^{2}}\nonumber\\
   & \approx  & \left\{\,2\,\left[\, \frac{(1+\tau)^{^{2}}}{4\,\tau}\,\,\Delta\epsilon \,+ \, \frac{\Delta y_{_{\rm GH}}}{{\rm w}^{\2}(z)}\,\,Y \,\right]\,\exp\left[-\,\frac{Y^{^{2}}}{{\rm w}^{\2}(z)}\,\right]\,\right\}^{^{2}}\,\,.
\end{eqnarray}
Finally, noting that, in the critical region, $(1+\tau)^{^{2}}\approx 4\,\tau$ (see Figs.\,5b-d), we can  get a further simplification of the outgoing beam intensity,
\begin{equation}
\label{eqsim}
I_{_{\rm out}}(Y,\Delta \epsilon) \, \propto \,
  \left[\, \Delta\epsilon \,+ \, \frac{\Delta y_{_{\rm GH}}}{{\rm w}^{\2}(z)}\,\,Y \,\right]^{^{2}}\,\exp\left[-\,\frac{2\,Y^{^{2}}}{{\rm w}^{\2}(z)}\,\right]\,\,.
\end{equation}

\section*{\large \color{navy} IV. THE PEAKS BEHAVIOR IN THE CRITICAL ANGLE REGION}

The starting point in optical weak measurement experiments is to set the angles of the first and second polarizers to
\[ \{\,\alpha_{\0}\,,\,\beta_{\0}\,\} = \{\,\mbox{$\frac{\pi}{4}$}\,,\,\mbox{$\frac{3\,\pi}{4}$}  + \gamma_{\0}\,\}\,\,.  \]
For this choice ($\Delta \epsilon = 0$) we find that the outgoing intensity,
\begin{equation}
I_{_{\rm out}}(Y,0) \, \propto \, Y^{^{2}}\,\exp\left[-\,\frac{2\,Y^{^{2}}}{{\rm w}^{\2}(z)}\,\right]\,\,,
\end{equation}
is a symmetric function with two peaks centered  at $Y_{_{\rm max}}^{^{\pm}} = \pm\,\,{\rm w}(z)/\sqrt{2}$ and a minimum centered at $Y_{_{\rm min}}=0$. By changing the angle of the second polarizer from $\beta_{\0}$ to $\beta_{\0}+\Delta \gamma$, we break the symmetry. In terms of $\Delta \epsilon=\Delta \gamma/\cos^{\2}\gamma_{\0}$,   we find
\begin{equation}
Y_{_{\rm min}}(\Delta \epsilon) = -\,\frac{\Delta \epsilon}{\Delta y_{_{\rm GH}}}\,\,{\rm w}^{\2}(z)
\end{equation}
and
\begin{equation}
\label{eqmax}
Y_{_{\rm max}}^{^{\pm}}(\Delta \epsilon) = \frac{-\,\Delta \epsilon\, \pm
\sqrt{(\Delta \epsilon)^{^{2}}+ 2\, [\,\Delta y_{_{\rm GH}}^{^{2}}/\,\,{\rm w}^{\2}(z)\,]   }}{2\, \Delta y_{_{\rm GH}}}\,\,{\rm w}^{\2}(z)\,\,.
\end{equation}
It is clear that for positive  $\Delta \epsilon$ (anti-clockwise rotation $\Delta \gamma$ around $\gamma_{\0}$), the $Y_{_{\rm max}}^{^{+}}(|\Delta \epsilon|)$ represents the position of  the main peak of the outgoing beam. For negative $\Delta \epsilon$, the main peak is instead centered at $Y_{_{\rm max}}^{^{-}}(-|\Delta \epsilon|)$.
By using  Eq.\,(\ref{eqmax}), the distance between these peaks is given by
\begin{equation}
\label{for1}
\Delta Y_{_{\rm max}} = Y_{_{\rm max}}^{^{+}}(|\Delta \epsilon|) - Y_{_{\rm max}}^{^{-}} (-|\Delta\epsilon|)
 = \frac{-\,|\Delta \epsilon|\,+\,
\sqrt{|\Delta \epsilon|^{^{2}}+ 2\, [\,\Delta y_{_{\rm GH}}^{^{2}}/\,\,{\rm w}^{\2}(z)\,]   }}{\Delta y_{_{\rm GH}}}\,\,{\rm w}^{\2}(z)\,\,.
 \end{equation}
In the region $0\leq  |\Delta \epsilon| \leq \Delta y_{_{\rm GH}}/{\rm w}(z)$, we find
 \begin{equation}
\sqrt{2}\,\,{\rm w}(z)\,\,\leq  \,\,\Delta Y_{_{\rm max}} \,\leq \,\,(\sqrt{3}-1)\,\,{\rm w}(z)\,\,.
\end{equation}
This clearly shows that by increasing the value of $|\Delta \epsilon|$,  we reduce the distance between the peaks.      For $|\Delta \epsilon| \gg \Delta y_{_{\rm GH}}/{\rm w}(z)$,
\begin{equation}
\label{epsgg}
 \Delta Y_{_{\rm max}}\, \approx \,\,\Delta y_{_{\rm GH}}/|\Delta \epsilon|\,\,.
\end{equation}
For incidence angle far from the critical region, due to the fact that the GH shift is proportional to the wavelength of the laser beam, the condition
\[ \frac{\Delta y_{_{\rm GH}}}{{\rm w}(z)}\,\, \approx \,\frac{\lambda}{{\rm w}(z)}\,\ll\, |\Delta \epsilon| \]
can be easily satisfied. Thus, far from the critical region, the experimental curves of $\Delta Y_{_{\rm max}}$ reproduce the GH curves amplified by the factor $1/|\Delta \epsilon|$.

In the critical region, the frequency crossover and the axial dependence, showed in Fig.\,3 (BK7) and Fig.\,4 (fused silica), play against the validity of the constraint $|\Delta \epsilon|\gg \Delta y_{_{\rm GH}}/{\rm w}(z)$. This means that in this region, the experimental curves  of $\Delta Y_{_{\rm max}}$ do not necessarily reproduce the GH curves. The expected experimental curves of the peaks distance are plotted for different value of  $|\Delta \epsilon|$ in Fig.\,6 (BK7) and Fig.\,7 (fused silica).  The plots confirm that, at critical incidence, the amplification does {\em not} reproduce  the $1/|\Delta \epsilon|$ proportionality. The axial dependence is removed by increasing the beam waist ${\rm w}_{\0}$.

For experimental use, it is convenient to express the GH shift, $\Delta y_{_{\rm GH}}$,  in terms of the experimental  quantity $\Delta Y_{_{\rm max}}$. From Eq.\,(\ref{for1}), we obtain
\begin{equation}
\label{for2}
\Delta y_{_{\rm GH}} = \frac{2\,|\Delta \epsilon|\,{\rm w}^{\2}(z)}{2\,{\rm w}^{\2}(z) - \Delta Y^{^{2}}_{_{\rm max}}}\,\,\Delta Y_{_{\rm max}}\,\,.
\end{equation}
The error on the GH shift is given by
\[
\frac{\sigma(\Delta y_{_{\rm GH}})}{\Delta y_{_{\rm GH}}} =
\sqrt{\left[\,\frac{\sigma(|\Delta \epsilon|)}{|\Delta \epsilon|}\,\right]^{^{2}} +\,\, \left[\,
\frac{2\,{\rm w}^{\2}(z) + \Delta Y^{^{2}}_{_{\rm max}}}{2\,{\rm w}^{\2}(z) - \Delta Y^{^{2}}_{_{\rm max}}}\,\,\frac{\sigma(\Delta Y_{_{\rm max}} )}{\Delta Y_{_{\rm max}}}
 \right]^{^{2}}   + \,\, \left\{\,
\frac{2\,\Delta Y^{^{2}}_{_{\rm max}}}{2\,{\rm w}^{\2}(z) - \Delta Y^{^{2}}_{_{\rm max}}}\,\,\frac{\sigma[{\rm w}(z) ]}{ {\rm w}(z)  }
 \right\}^{^{2}}      }  \,\,.
\]
Recalling that for $\Delta \epsilon=0$, the distance between the peaks gives a direct information on the beam waist,
$\Delta Y_{_{\rm max}}=\sqrt{2}\,{\rm w}(z)$. we  can use
\[ \sigma(\Delta Y_{_{\rm max}} ) = \sigma[{\rm w}(z)] \]
in the previous error formula and obtain
\[
\frac{\sigma(\Delta y_{_{\rm GH}})}{\Delta y_{_{\rm GH}}} =
\sqrt{\left[\,\frac{\sigma(|\Delta \epsilon|)}{|\Delta \epsilon|}\,\right]^{^{2}} +
\,\,\left[\,
\frac{2\,{\rm w}^{\2}(z) + \Delta Y^{^{2}}_{_{\rm max}}}{2\,{\rm w}^{\2}(z) - \Delta Y^{^{2}}_{_{\rm max}}}
 \right]^{^{2}}
  \left\{ 1   +  \left[\,\frac{2\,\Delta Y^{^{3}}_{_{\rm max}}/\,{\rm w}(z)}{2\,{\rm w}^{\2}(z) + \Delta Y^{^{2}}_{_{\rm max}}}  \right]^{^{2}}    \right\}
 \left[\, \frac{\sigma(\Delta Y_{_{\rm max}} )}{\Delta Y_{_{\rm max}}}\,
 \right]^{^{2}}     }  \,\,.
\]
In the region  ${\rm w}(z) \leq  \Delta Y_{_{\rm max}}  \leq  \sqrt{2}\,{\rm w}(z)$, we find
\[
13 \leq  \left[\,
\frac{2\,{\rm w}^{\2}(z) + \Delta Y^{^{2}}_{_{\rm max}}}{2\,{\rm w}^{\2}(z) - \Delta Y^{^{2}}_{_{\rm max}}}
 \right]^{^{2}}
  \left\{ 1+
  \left[\,\frac{2\,\Delta Y^{^{3}}_{_{\rm max}}/\,{\rm w}(z)}{2\,{\rm w}^{\2}(z) + \Delta Y^{^{2}}_{_{\rm max}}}  \right]^{^{2}}    \right\}\leq \infty\,\,.
\]
To avoid great standard deviations, we have to work in the region  $\Delta Y_{_{\rm max}} \leq  {\rm w}(z)$,
 where
\begin{equation}
\sqrt{\left[\,\frac{\sigma(|\Delta \epsilon|)}{|\Delta \epsilon|}\,\right]^{^{2}} +
\,\,
 \left[\, \frac{\sigma(\Delta Y_{_{\rm max}} )}{\Delta Y_{_{\rm max}}}\,
 \right]^{^{2}}     }\,\,\leq\,\, \frac{\sigma(\Delta y_{_{\rm GH}})}{\Delta y_{_{\rm GH}}} \,\,\leq \,\,
\sqrt{\left[\,\frac{\sigma(|\Delta \epsilon|)}{|\Delta \epsilon|}\,\right]^{^{2}} +
\,\,13\,
 \left[\, \frac{\sigma(\Delta Y_{_{\rm max}} )}{\Delta Y_{_{\rm max}}}\,
 \right]^{^{2}}     }  \,\,.
\end{equation}
From the previous condition on $\Delta Y_{_{\rm max}}$, by using Eq.(\ref{for2}) we obtain
\begin{equation}
\label{min}
|\Delta \epsilon|\,\, \geq \,\,\frac{\Delta y_{_{\rm GH}}}{2\,{\rm w}(z)}\,\,.
\end{equation}
The choice of $|\Delta \epsilon_{_{\rm min}}| = \Delta y_{_{\rm GH}}/2\,{\rm w}(z)$ in the second polarizer thus represents an additional experimental constraint to avoid great standard deviations.

\section*{\large \color{navy} V. CONCLUSIONS AND OUTLOOKS}

The possibility to use the  weak measurement of the electron spin component\cite{R9}  in optics\cite{R8a,R8b} have recently stimulated the realization of an experiment\cite{R7},  based on the interference between different polarizations, in which the GH shift curves are reproduced in the region of the validity of the standard analytic formula (\ref{anaGH}). Nevertheless,  the analytical shift  (\ref{anaGH})
diverges when the incidence angle approaches the critical angle.  In a recent paper\cite{R5}, this divergence was removed and  an analytic formula, valid for $2\,z\ll k{\rm w}_{\0}^{\2}$, proposed for  the GH shift at critical angle,
\begin{equation}
\label{GHcri}
\left\{\,y^{^{[s]}}_{_{\rm GH}}\,,\,y^{^{[p]}}_{_{\rm GH}}\,\right\}_{_{\rm cri}}  \approx
\frac{\sqrt{k\,{\rm w}(z)}}{k}\,\sqrt{
 2\,\sqrt{2\,\pi}\,\,\,\frac{\sqrt{2 - n^{\2} + 2\, \sqrt{n^{\2}-1}}}{n^{^{2}}-1 + \sqrt{n^{^{2}}-1}}}\,\left\{\,1\,,\,n^{\2}\,\right\}\,\,.
\end{equation}
This closed formula, which is in excellent agreement with the numerical data plotted in Fig.\,3-c and Fig.\,4-c, clearly
shows the crossover frequency at critical angle. The amplification  $\sqrt{k\,{\rm w}(z)}$  at critical angle suggested studying with more care the peaks behavior  of the beam in optical weak measurements for incidence within the critical region. Indeed, in such a region, due to this amplification, the condition $|\Delta \epsilon|\gg \Delta y_{_{\rm GH}}/{\rm w}(z)$ and the consequent proportionality amongst the experimental curves of the peaks distance and the GH curves are no longer valid, see for example Fig.\,6-c and Fig.\,7-c.

In our study, we have also found an axial dependence  in optical weal measurements. This axial dependence can affect the experimental curves and,  for small second polarizer rotations and small values of the beam waist,
produces a practically flat region, see the plots in Fig.\,6-a and Fig.\,7-a for $|\Delta \epsilon|=0.01$. To minimize the axial dependence, we have to work with laser beam with ${\rm w}_{\0}\geq 500\,\mu{\rm m}$. It is important to be observed here that, also for ${\rm w}_{\0}=500\,\mu{\rm m}$, the curve amplification $1/|\Delta \epsilon|$  can be only reproduced far from the critical region.

In view of a possible experimental analysis of the study presented in this article, we have also estimated in which region we reach the better standard deviation for $\Delta y_{_{\rm GH}}$ in the critical region.  By using the second polarizer angle constraint,  Eq.(\ref{min}), and the analytical formula for the GH shift at critical angle, Eq.(\ref{GHcri}), we find
\begin{equation}
|\Delta \epsilon|\,\, \geq \,\,
\sqrt{
 \sqrt{\frac{\pi}{2}}\,\,\,\frac{\sqrt{2 - n^{\2} + 2\, \sqrt{n^{\2}-1}}}{n^{^{2}}-1 + \sqrt{n^{^{2}}-1}}}\,\,\,
 \frac{n^{\2} -1}{ \sqrt{k\,{\rm w}(z)} }  \,\,.
\end{equation}
This implies, for laser beams with ${\rm w}_{\0}= 500\,\mu{\rm m}$ and camera at $z\leq 50\,{\rm cm}$, $|\Delta \epsilon_{_{\rm min}}|\approx 0.015$ both for BK7 and fused silica dielectric blocks.

We conclude this work, by observing that our analysis  does not take into
account cumulative dissipations and imperfections in the dielectric prism (such as the misalignment of its surfaces) and the beam reshaping caused by interference . A phenomenological way to include misalignment effects is given in ref.\cite{del6}. An interesting discussion on the origin of negative and positive lateral shifts in a dielectric slab is investigated in ref.\cite{slab} form the viewpoint of the interference between multiple light beams.

\vspace{.8cm}

\noindent
\textbf{ACKNOWLEDGEMENTS}\\
The authors gratefully thank  the Capes (M.P.A.) and  CNPq (S.\,D.\,L. and G.\,G.\,M.) for the financial support. One of the authors (S.\,D.\,L.) is greatly indebted to Silv\^ania A. Carvalho for interesting comments and  stimulating discussions. Finally, the authors wish to thank the referees for their useful observations and suggestions.


\newpage

\begin{figure}
\vspace*{-1cm}
\hspace*{-1.5cm}
		\includegraphics[width=18.5cm, height=22.cm, angle=0]{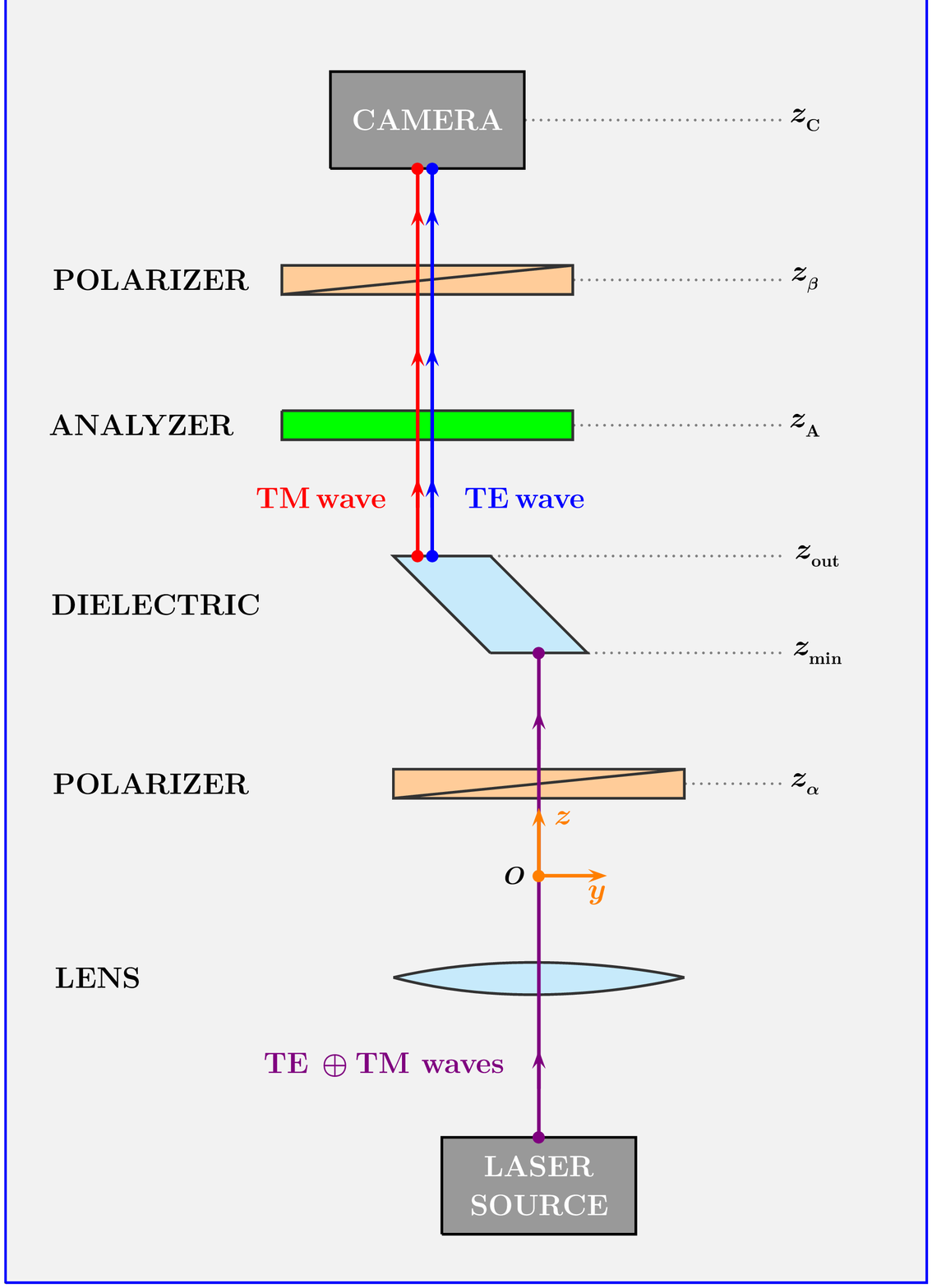}
\vspace*{-1.6cm}
\caption{{\bf Experimental layout.} A schematic representation of the optical weak measurement experiment for the observation of the transversal distance between the main peak of the outgoing beams for two opposite rotations, $\pm\Delta \gamma$, of the second polarizer.  The incoming beam, which, after the first polarizer $(\alpha=\pi/4)$,  has an equal mixture of $s$ and $p$ polarized waves, passes through the dielectric block ($z_{_{\rm in}}<z< z_{_{\rm out}}$) and then passing through the analyzer in $z=z_{_{\rm A}}$ loses the global $\Delta \phi_{_{\rm GH}}$ phase). Optical weak measurements are done by
changing the rotation angle in the second polarizer ($\beta =3\pi/4 +\gamma_{\0} \pm |\Delta \gamma|$).  The angle $\gamma_{\0}$ is fixed to obtain, for $\Delta \gamma =0$, an outgoing beam with two identical maxima centered at $\pm\, {\rm w}(z)/\sqrt{2}$.}
\end{figure}

\newpage

\begin{figure}
\vspace*{-5cm}
\hspace*{-1.7cm}
		\includegraphics[width=18.5cm, height=29.5cm, angle=0]{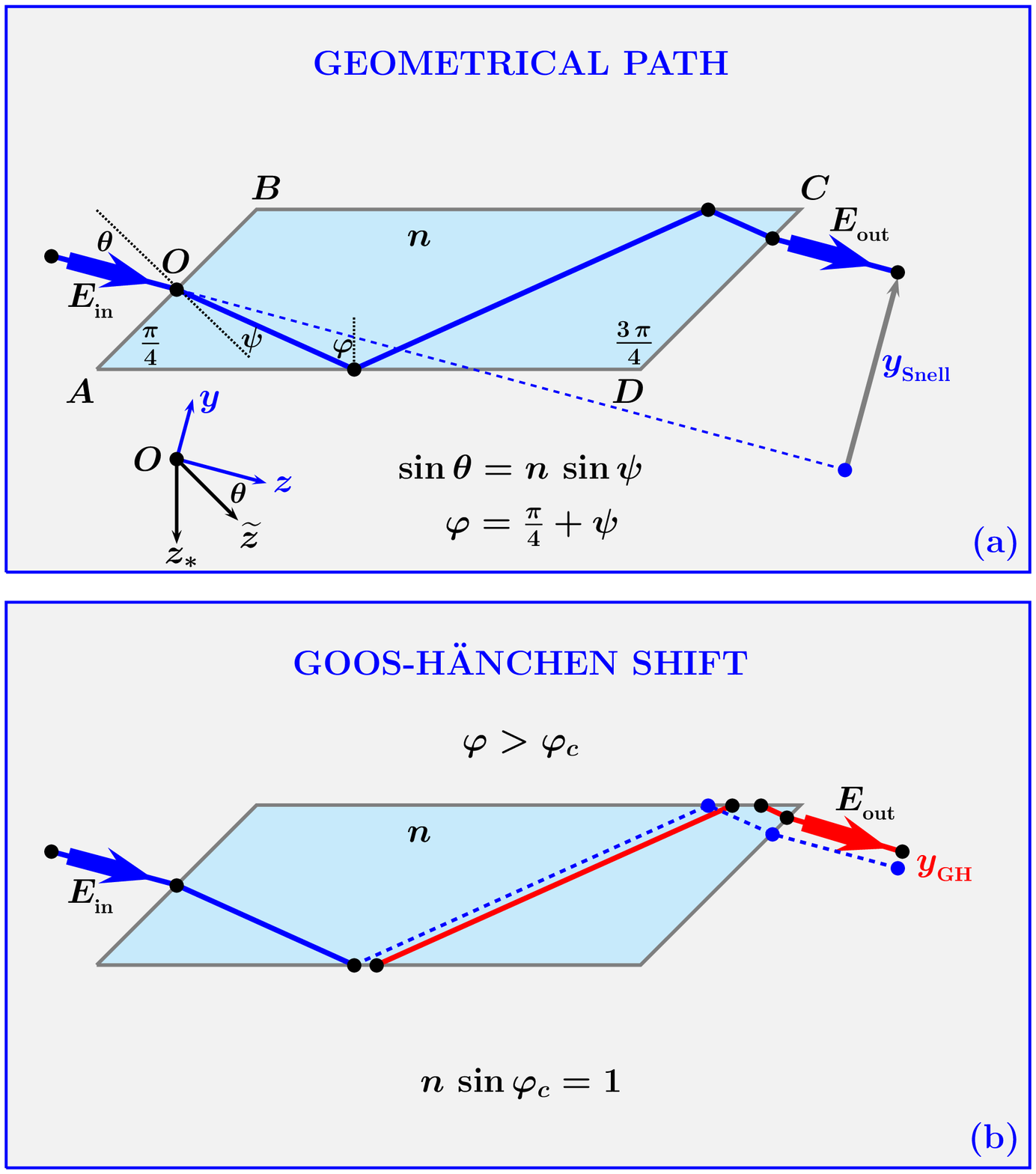}
\vspace*{-4.5cm}
\caption{{\bf Geometrical path and Goos-H\"anchen shift.} Schematic diagram of the dielectric block
analyzed in this paper. In (a), it is shown the geometrical path predicted by the Snell law, Eq.\,(\ref{SnellP}).
For $\varphi>\varphi_c$, an additional phase, coming from the Fresnel reflection coefficients at the down and up interfaces, has to be considered. This phase is responsible for the addition shift,
Eq.\,(\ref{anaGH}), known as Goos-H\"anchen shift and shown in (b).}
\end{figure}

\newpage

\begin{figure}
\vspace*{-2.5cm}
\hspace*{-2.2cm}
		\includegraphics[width=19cm, height=26.5cm, angle=0]{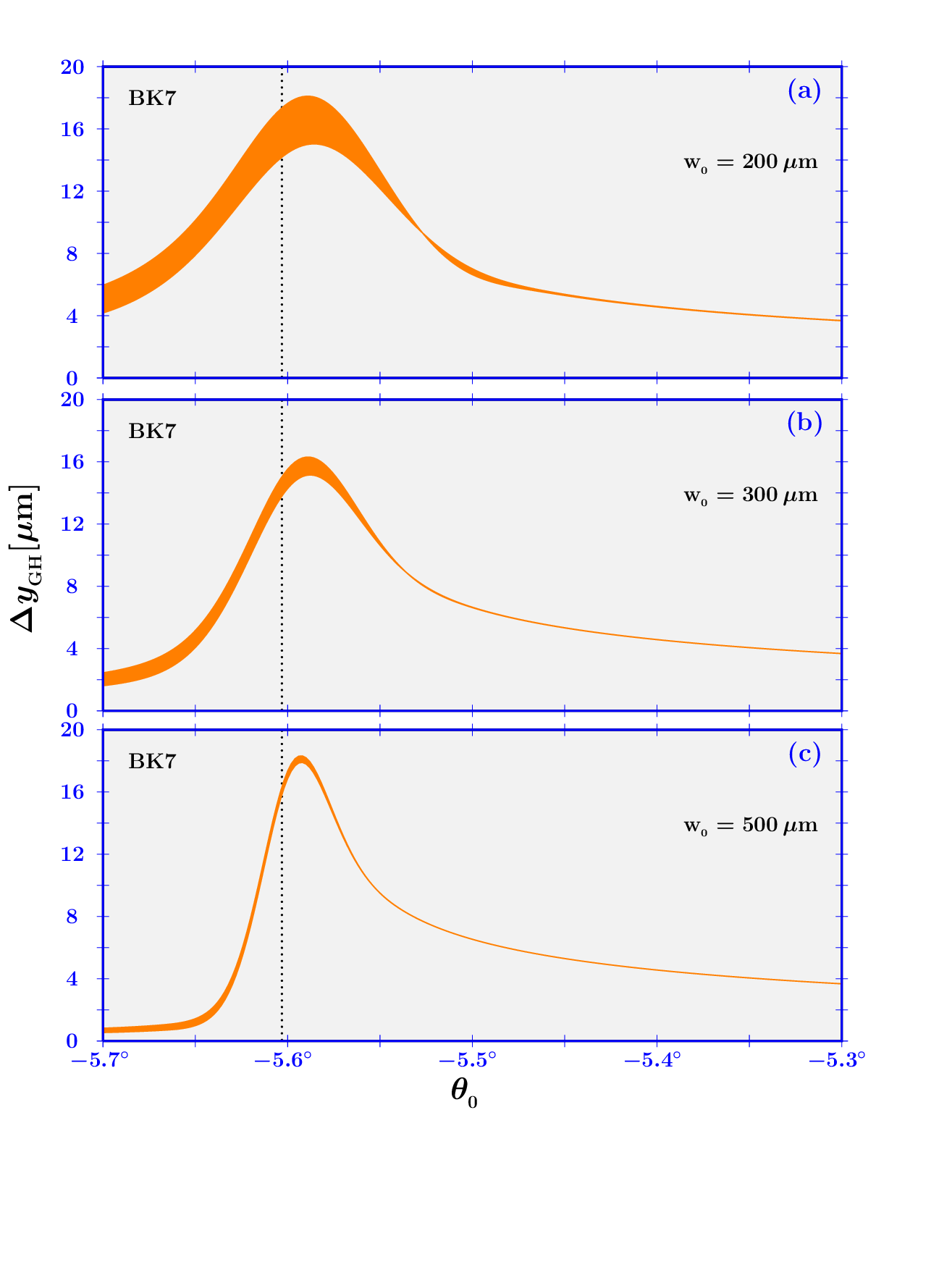}
\vspace*{-4.cm}
\caption{{\bf The GH shift curves for BK7 blocks.} The numerical data for the transversal $\Delta y_{_{\rm GH}}$  shift of laser gaussian beams,  passing through a BK7 dielectric block, are plotted, in the axial range $10\,{\rm cm} \leq z \leq 15\,{\rm cm}$ for different beam waists ${\rm w}_{\0}=200\,\mu{\rm m}$ (a), $300\,\mu{\rm m}$ (b), and
 $500\,\mu{\rm m}$ (c). The crossover frequency at critical angle is clear from the plots and the axial dependence represents  an additional  phenomenon to be considered in optical weak measurements.}
\end{figure}

\newpage

\begin{figure}
\vspace*{-2.5cm}
\hspace*{-2.2cm}
		\includegraphics[width=19cm, height=26.5cm, angle=0]{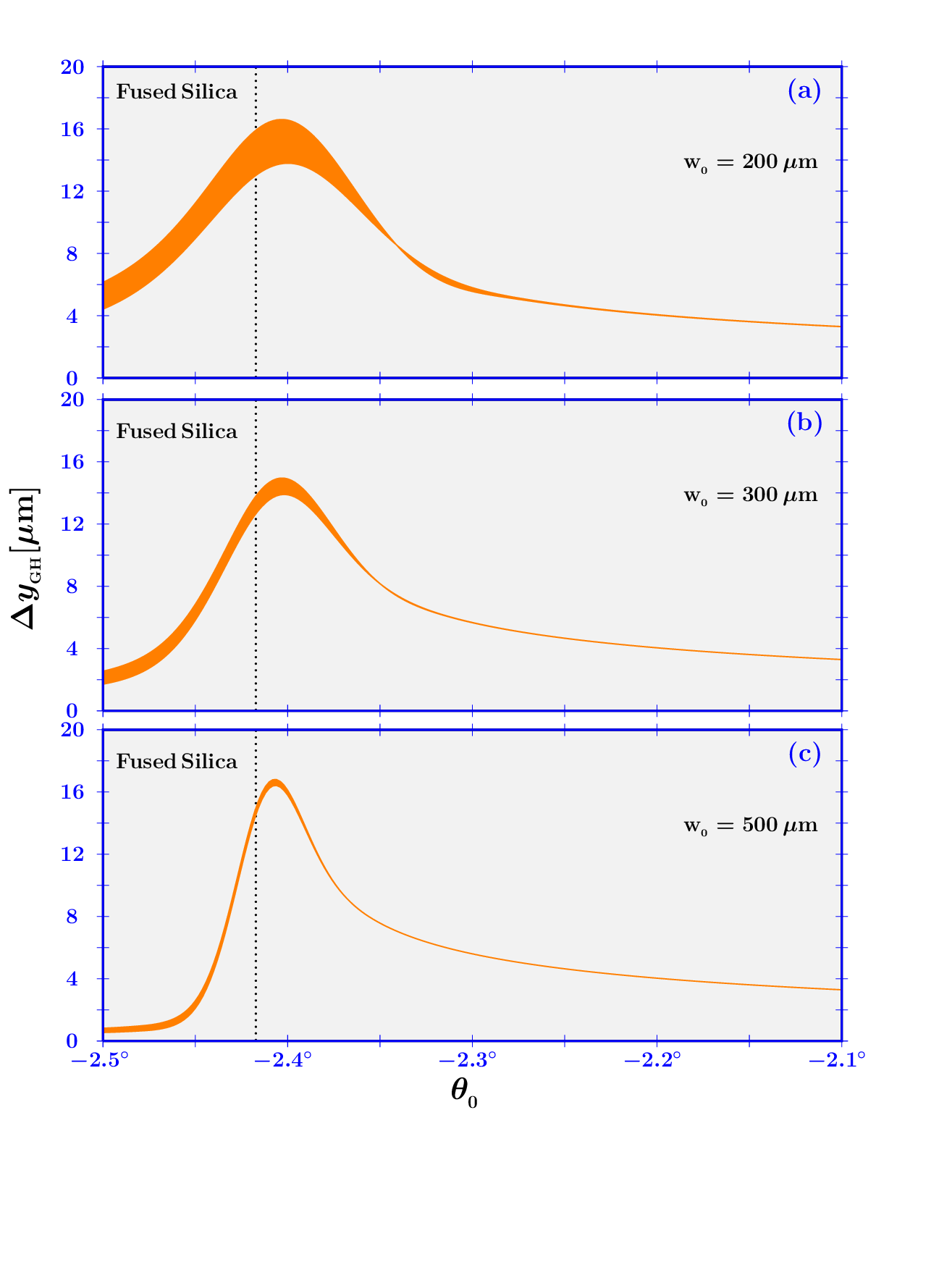}
\vspace*{-4.cm}
\caption{{\bf The GH shift curves for fused silica blocks.} The numerical data for the transversal $\Delta y_{_{\rm GH}}$  shift of laser gaussian beams,  passing through a fused  dielectric block, are plotted, in the axial range $10\,{\rm cm} \leq z \leq 15\,{\rm cm}$, for different beam waists ${\rm w}_{\0}=200\,\mu{\rm m}$ (a), $300\,\mu{\rm m}$ (b), and
 $500\,\mu{\rm m}$ (c). The crossover frequency at critical angle is clear from the plots and the axial dependence represents  an additional  phenomenon to be considered in optical weak measurements.}
\end{figure}

\newpage

\begin{figure}
\vspace*{-2.5cm}
\hspace*{-2.2cm}
		\includegraphics[width=19cm, height=25.8cm, angle=0]{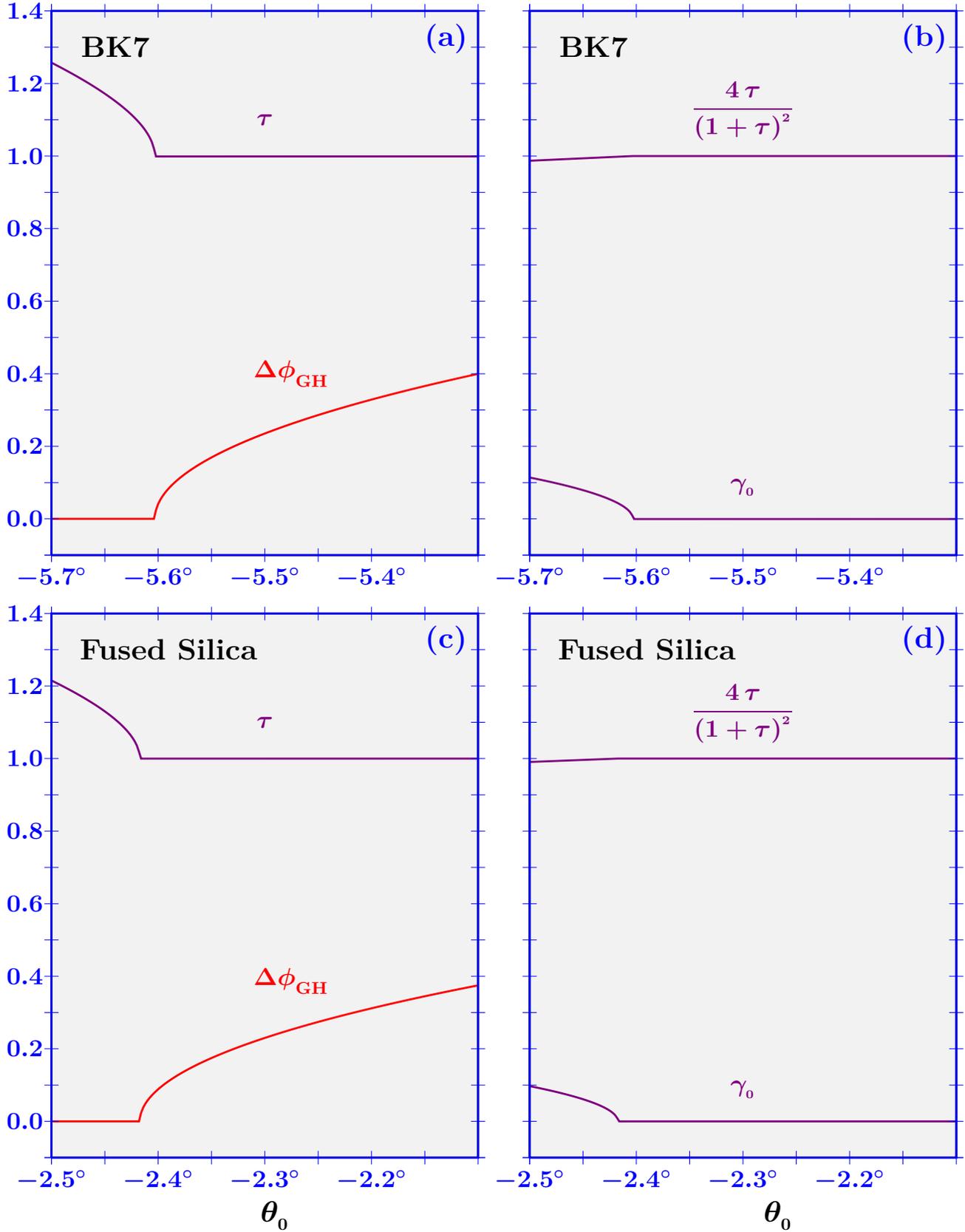}
\vspace*{-3.4cm}
\caption{{\bf Angular dependence of $\boldsymbol{\tau}$,  $\boldsymbol{\Delta \phi_{_{\rm GH}}}$, and  $\boldsymbol{\gamma_{\0}}$.} The angular dependence of $\tau$ (ratio between the modulus of the amplitudes for $s$ and $p$ polarized light)  and  $\Delta \phi_{_{\rm GH}}$  (global  phase difference between $s$ and $p$  polarized waves) are plotted  in (a) for BK7 and (c) for fused silica blocks. The fact that, in the critical region,  $4\,\tau/(1+\tau)^{^{2}}$ is practically equal to one, it is very useful to simplify the expression for the outgoing beam, see  Eq.\,(\ref{eqsim}). The numerical data for $\gamma_{\0}$ permits to calculate  the second polarizer angle ($\beta_{\0}=3\pi/4+\gamma_{\0}$) for which we find  an outgoing beam with two identical maxima centered at $\pm\,{\rm w}(z)$.}
\end{figure}

\newpage

\begin{figure}
\vspace*{-2.5cm}
\hspace*{-2.2cm}
		\includegraphics[width=19cm, height=25.8cm, angle=0]{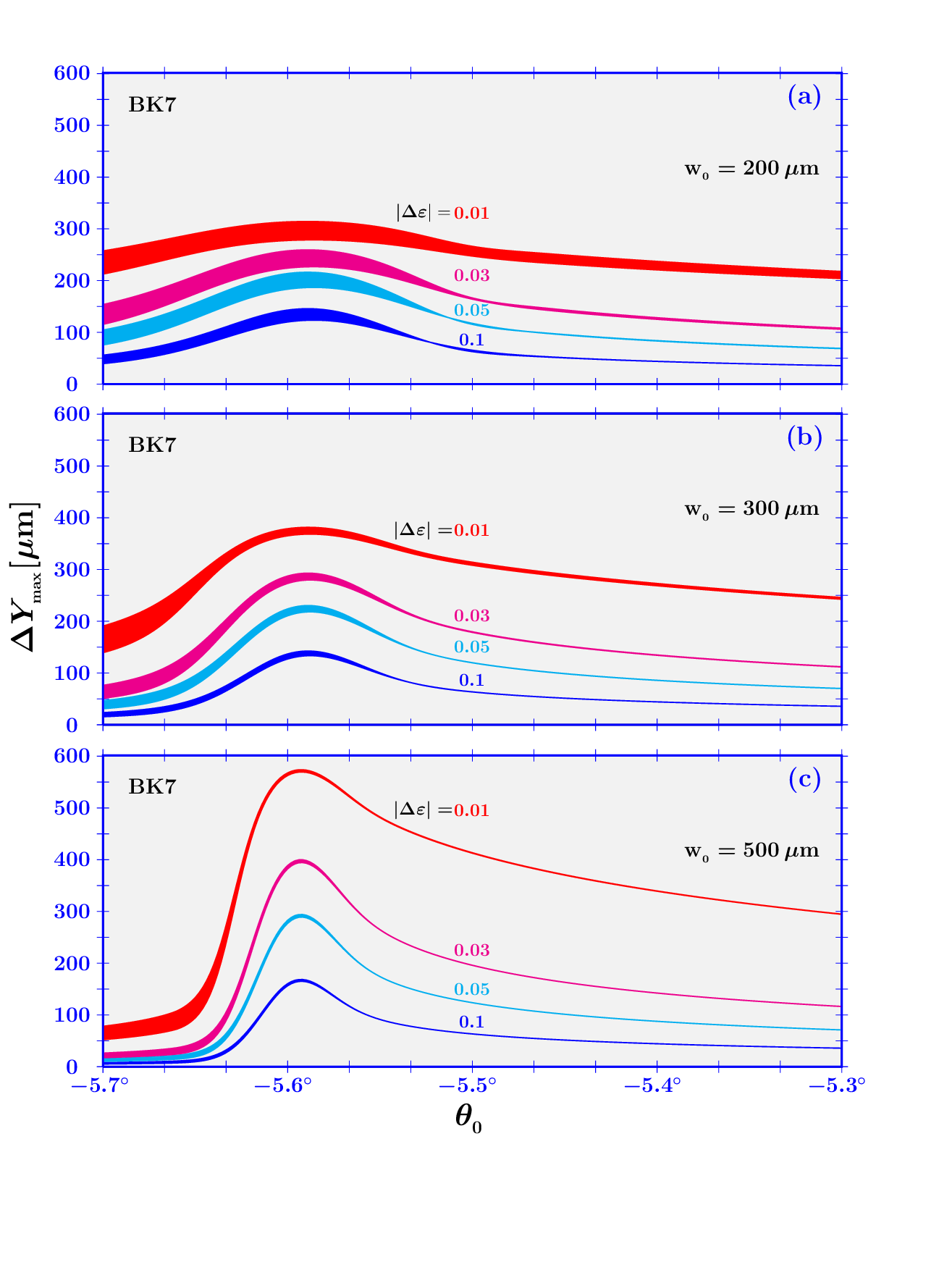}
\vspace*{-3.6cm}
\caption{{\bf Optical weak measurements curves for BK7 blocks.} The expected curves for the distance between the main peak of the beams coming out from a BK7 dielectric block and passing through the second polarizer for two opposite rotations, $|\Delta \epsilon|=|\Delta \gamma|/\cos^{\2}\gamma_{\0}$, are plotted   in the axial range $10\,{\rm cm} \leq z \leq 15\,{\rm cm}$, for different beam waists ${\rm w}_{\0}=200\,\mu{\rm m}$ (a), $300\,\mu{\rm m}$ (b), and
 $500\,\mu{\rm m}$ (c). From the plots, it is clear that to improve the crossover frequency and to reduce the axial dependence,  we have to work with ${\rm w}_{\0}\geq 500\,\mu{\rm m}$. Note that, also working with ${\rm w}_{\0}= 500\,\mu{\rm m}$, the curve amplification $1/|\Delta \epsilon|$, valid for incidence far from the critical region, is lost when the incidence angle approaches the critical one. }
\end{figure}

\newpage

\begin{figure}
\vspace*{-2.5cm}
\hspace*{-2.2cm}
		\includegraphics[width=19cm, height=25.8cm, angle=0]{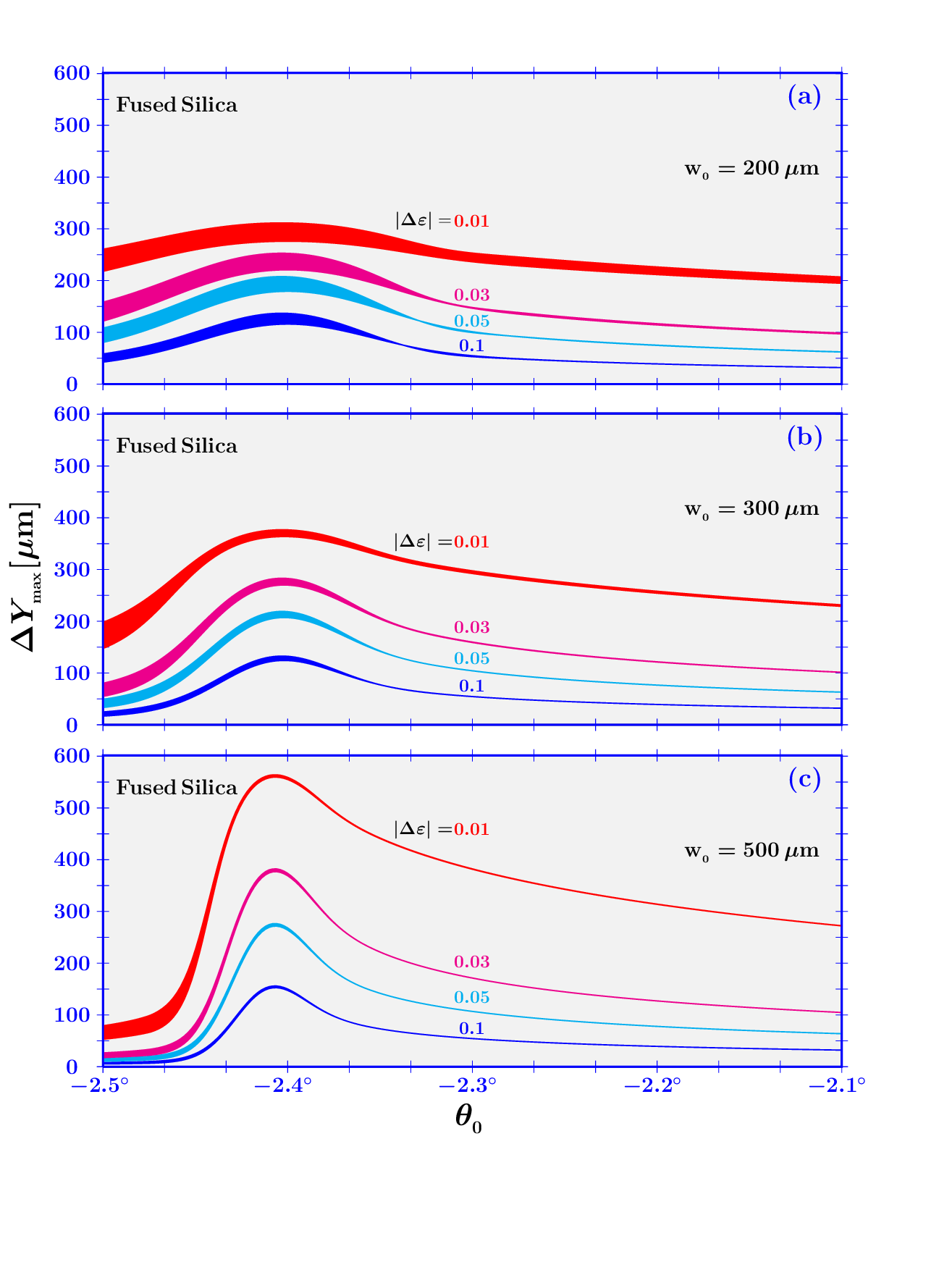}
\vspace*{-3.6cm}
\caption{{\bf Optical weak measurements curves for fused silica blocks.} The expected curves for the distance between the main peak of the beams coming out from a fused silica dielectric block and passing through the second polarizer for two opposite rotations, $|\Delta \epsilon|=|\Delta \gamma|/\cos^{\2}\gamma_{\0}$, are plotted   in the axial range $10\,{\rm cm} \leq z \leq 15\,{\rm cm}$, for different beam waists ${\rm w}_{\0}=200\,\mu{\rm m}$ (a), $300\,\mu{\rm m}$ (b), and
 $500\,\mu{\rm m}$ (c). From the plots, it is clear that to improve the crossover frequency and to reduce the axial dependence,  we have to work with ${\rm w}_{\0}\geq 500\,\mu{\rm m}$. Note that, also working with ${\rm w}_{\0}= 500\,\mu{\rm m}$, the curve amplification $1/|\Delta \epsilon|$, valid for incidence far from the critical region, is lost when the incidence angle approaches the critical one. }
\end{figure}

\end{document}